\documentclass[aps,prl,twocolumn,nopacs,groupedaddress]{revtex4-1}
\usepackage{amsfonts}
\usepackage{amsmath}
\usepackage{amssymb}
\usepackage{graphicx}
\usepackage{bm}
\usepackage{xcolor}

\renewcommand{\d}{\text{d}}

\begin{document}

\title{Effect of surface-active contaminants on radial  thermocapillary flows}
\author{T. Bickel}
\email[E-mail: ]{thomas.bickel@u-bordeaux.fr}
\affiliation{Univ. Bordeaux, CNRS, Laboratoire Ondes et Mati\`ere d'Aquitaine (UMR 5798), F-33400 Talence, France}

\begin{abstract}
We study the thermocapillary creeping flow induced by a thermal gradient at the liquid-air interface in the presence of insoluble  surfactants (impurities). Convective sweeping of the surfactants  causes density inhomogeneities that confers  in-plane elastic features to the interface. This mechanism is discussed for radially symmetric temperature fields, in both the deep and shallow water regimes. When mass transport is controlled by convection, it is found that  surfactants are depleted from a region whose size is inversely proportional to the interfacial elasticity.  Both the concentration and the velocity fields follow power laws at the border of the depleted region. Finally, it is shown that this singular behavior is smeared out when molecular diffusion is accounted for.
\end{abstract}

\maketitle

\section{Introduction}

When an interface is exposed to a temperature gradient, the local variation of its surface tension induces a shear stress that drives the fluids in motion~\cite{scrivenNature1960}. The resulting thermocapillary (or Marangoni) flows are ubiquitous in everyday life as well as in industrial processes. Their structure and stability have been thoroughly documented  in the fluid mechanics literature~\cite{levichbook,davisARFM1987}. For instance,  it has been known since the $18^{th}$ century that thermocapillarity can be used to actuate floating bodies~\cite{vandermens1869}. Another example is provided by the B\'enard-Marangoni instability: when a thin liquid layer is heated from below, it exhibits a regular flow pattern consisting of hexagonal cells that are attributed to thermocapillary convection~\cite{sefianeACIS2007}.

Current scientific challenges involving thermocapillary flows are focused on microscopic scales~\cite{karbalaeiMM2016}. Indeed, the prevalence of interfacial phenomena as the size decreases allows for a fine control of free surface flows. A convenient way to  remotely supply heat at  microscales is provided by light absorption from a focused laser beam. Thermocapillarity allows for instance the formation of vortical flows at low Reynolds number, with possible applications for mixing~\cite{muruJACS2006}. The production, transport and fusion of droplets  have been achieved in microfluidic devices using the photothermal effect~\cite{baroudLC2007}. Likewise, it has been shown that the height profile of nanometer-thin films can be controlled by light-induced temperature gradients~\cite{pottierPRL2015}.  Direct conversion of light into work has also been evidenced  in the field of active matter:  actuation of  microfabricated rotors~\cite{maggiNatCom2015} and micron-sized spherical particles~\cite{girotLangmuir2016,zhongOE2017} at the liquid-air interface have been accomplished recently, confirming that thermocapillarity is a highly efficient mechanism for colloidal self-propulsion.

From a theoretical viewpoint, the structure of  thermocapillary flows is fairly well understood~\cite{bratukhin1967,birikhbook}. 
Still, a body of experimental evidence reveals that the situation is not always as clear as expected. For instance, horizontal velocities measured along the interface can be several orders of magnitude lower than that predicted for Marangoni convection~\cite{itoJSME1990,wuIchmt2000,wuIchmt2001}. The divergent flows that originate from a radially symmetric source may also be unstable with respect to azimuthal perturbations, both at the macroscopic~\cite{favrePoF1997,mizevPoF2005,shternJFM1993} and the microscopic~\cite{girotLangmuir2016} scales. To account for these observations, contamination of the liquid-air interface by surface-active species is commonly invoked. Indeed, the convective sweeping of surfactants by the thermally-induced flow causes their accumulation at the boundaries of the container. The ensuing tension gradient --- whose origin is now solutal ---  then gives rise to restoring Marangoni forces and confers an effective in-plane elasticity to the interface. This mechanism was first discussed by Berg and Acrivos in the context of the B\'enard-Marangoni instability~\cite{bergCES1965}. It is actually reminiscent of Levich's explanation for the retarded motion of ascending bubbles in a liquid~\cite{levichbook,ybertEPJE2000}.

In the present study, we consider the hydrodynamic response of a surfactant-laden interface to a thermocapillary flow. Although many experiments are performed with radially symmetric sources~\cite{girotLangmuir2016,wuIchmt2000,wuIchmt2001,favrePoF1997,mizevPoF2005}, little is known regarding  3D axisymmetric geometry. Also, arbitrary liquid depths have not yet received much consideration. 
We thus intend to pursue previous theoretical works, that were mainly focused on  2D configurations in  the  shallow-water limit~\cite{homsyJFM1984,carpenterJFM1985,shmyrovAcis2018}. 
Here, the distribution of insoluble surfactants is  characterized in both  deep and shallow water configurations.
We also revisit the classical surface stagnation hypothesis of Carpenter and Homsy~\cite{carpenterJFM1985} and extend the analysis to mass transport regimes where both convection and diffusion are relevant.
The paper is organized as follows. In sect.~\ref{formulation}, we present the theoretical model and comment the main assumptions. The problem is then solved in sects.~\ref{deep} and~\ref{shallow} for various regimes of mass transport, depending on whether surface diffusion or convection --- or both --- are relevant. The main outcomes are finally summarized and discussed in sect.~\ref{conclusion}.

\section{Formulation of the problem}
\label{formulation}

\begin{figure}
\centering
\includegraphics[width=0.9\columnwidth]{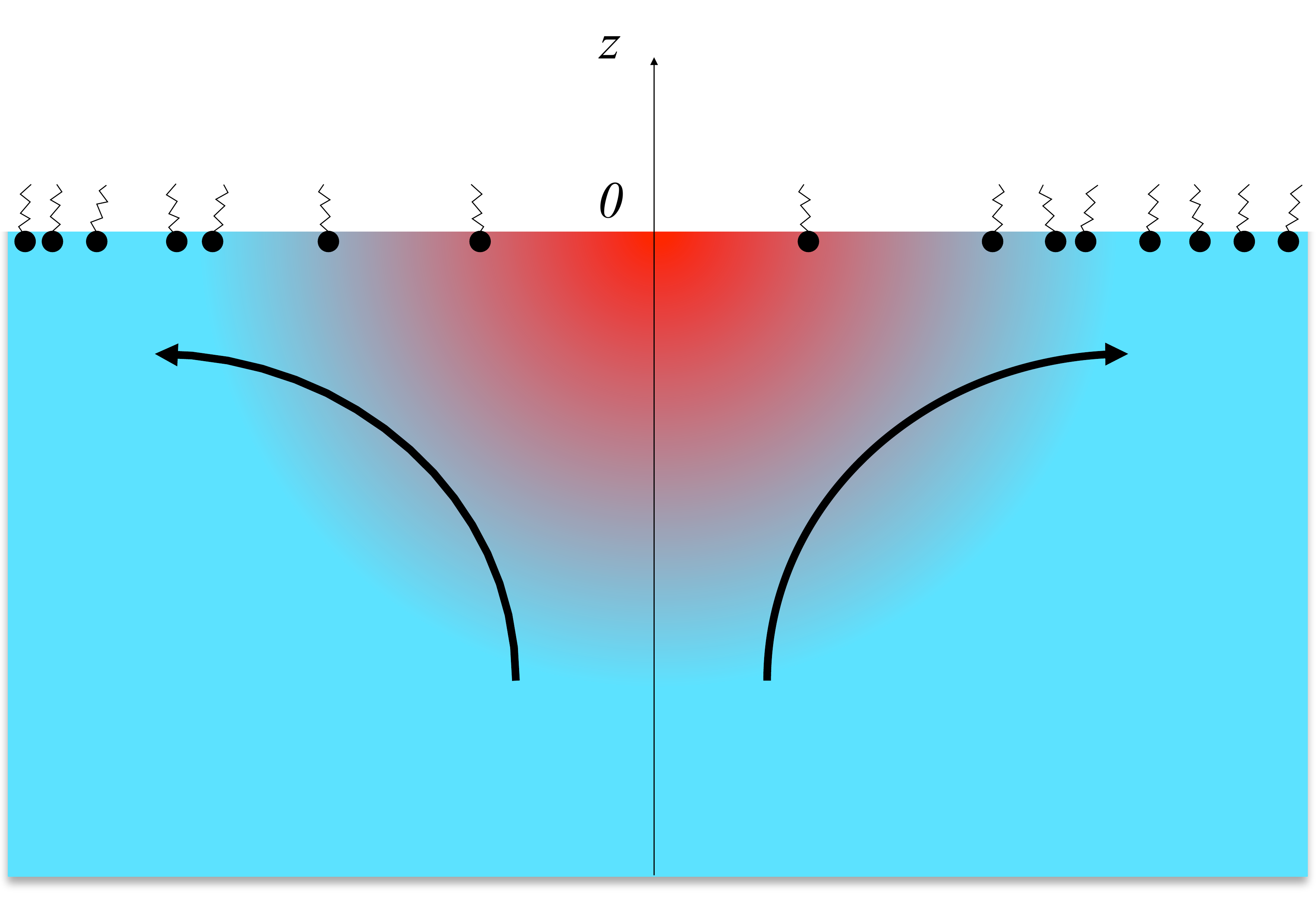}
\caption{Thermocapillary flow driven by a radially symmetric heat source, in the presence of surface-active impurities (represented as surfactants). The temperature gradient induces horizontal stresses that drive the liquid from the hot to the cold regions of the interface. Impurities are then swept away toward the cell boundaries, which induces a  gradient of concentration. }
\label{schema}       
\end{figure}

 The situation under investigation is schematically drawn in fig.~\ref{schema}. A newtonian, incompressible liquid of viscosity~$\eta$  and  mass density$\rho$   is enclosed in a cylindrical cell of radius~$R$ and height~$H$. The $z$-axis coincides with the centerline of the cell. It is oriented upwards, with unit vector $\mathbf{e}_z$. The free liquid-air interface is assumed to  remain flat and located at $z=0$. 
We denote $\mathbf{A}_{\parallel} = (\mathbf{1} - \mathbf{e}_z\mathbf{e}_z)\cdot \mathbf{A}$ the  horizontal projection of the vector~$\mathbf{A}$.

A small amount of bulk-insoluble surfactants is present at the liquid-gas interface. The equilibrium surface concentration~$\Gamma_0$ is assumed to be sufficiently small such that the surfactants remain in the dilute phase. Under non-uniform heating, the liquid is driven into motion by the thermocapillary effect. The resulting inhomogeneities in surfactant concentration then give rise to a solutal counterflow, that we aim to characterize.

\subsection{Marangoni flow}

The coupling between thermal and solutal Marangoni flows requires to solve simultaneously the momentum, energy and mass conservation equations~\cite{khattariJPCM2002,dasPoF2017,srivasEPJE2018}.
The flow velocity $\mathbf{v}(\mathbf{r},t)$ and pressure $p(\mathbf{r},t)$  are governed by the  Navier-Stokes equation
\begin{equation}
\rho \left( \partial_t \mathbf{v} + \mathbf{v} \cdot \bm{\nabla} \mathbf{v} \right)= \eta \nabla^2 \mathbf{v} - \bm{\nabla} p \ ,
\label{navierstokes}
\end{equation}
together with the continuity equation
\begin{equation}
\bm{\nabla} \cdot  \mathbf{v}  =0 \ .
\label{incomp}
\end{equation}
These equations are supplemented with the  boundary conditions at the free surface  
\begin{subequations}
\label{flowbc}
\begin{align}
& v_z\big\vert_{z=0} = 0 \ , \label{vz0} \\
& \eta \left( \partial_z \mathbf{v}_{\parallel} + \bm{\nabla}_{\parallel} v_z\right) \Big\vert_{z=0} = \bm{\nabla}_{\parallel} \gamma \ .
\label{bcmarangoni} 
\end{align}
\end{subequations}
The first condition eq.~(\ref{vz0}) expresses the absence of mass transport (\textit{e.g.}, evaporation) across the flat interface. The stress balance condition eq.~(\ref{bcmarangoni}) is the Marangoni boundary condition: it states that a fluid flow develops in the bulk as a result of  surface tension gradients.

Inhomogeneities of surface tension can be induced for instance by a thermal gradient. The surface tension $\gamma$ being a decreasing function of the temperature, the liquid is pulled from the hot to the cold regions of the interface. Throughout this work, it is assumed that
the liquid is heated by a radially symmetric heat source, with $q(\mathbf{r})$  the injected power density. The temperature field $T(\mathbf{r},t)$  is solution of the heat transfer equation
\begin{equation}
\rho c \left( \partial_t T + \mathbf{v} \cdot \bm{\nabla} T \right) = \kappa \nabla^2 T + q(\mathbf{r}) \ ,
\label{heatequation}
\end{equation}
where $c$ and $\kappa$ respectively stand for the heat capacity and thermal conductivity of the liquid. One also defines the thermal diffusivity  $\chi=\kappa/(\rho c)$.
Far from the heat source, the temperature is expected to return to its bulk value
\begin{equation}
\lim_{\vert \mathbf{r} \vert  \to \infty} T(\mathbf{r}, t) = T_0 \ . 
\label{tbulk}
\end{equation}
It is also assumed that there is no heat flux across the liquid-air interface
\begin{equation}
\partial_z T  \big\vert_{z=0} = 0 \ . 
\label{bcflux}
\end{equation}
The set of coupled eqs.~(\ref{navierstokes})--(\ref{bcflux}) defines the thermocapillary flow problem for a pure interface.

The situation gets more involved when a small amount of surface-active molecules are irreversibly adsorbed at the interface. 
Indeed, in addition to Fickian diffusion, the surfactants are also advected by the thermocapillary flow. 
The surface concentration $\Gamma (\mathbf{r}_{\parallel},t)$ of insoluble surfactants satisfies the transport equation
\begin{equation}
\partial_t \Gamma + \bm{\nabla}_{\parallel} \cdot \left( \mathbf{v}_{\parallel} \Gamma\right) = D \nabla_{\parallel}^2 \Gamma \ ,
\label{diffusion}
\end{equation}
with  $D$ the  diffusion coefficient along the interface. 
This equation is solved together with the no-flux condition at the boundary of the cell ($r=R$).

In the presence of surfactants, the  surface tension depends on both the temperature and  the concentration. A key ingredient of the analysis is  provided by the  equation of state that relates the surface tension $\gamma$ to the temperature~$T$ and the surface concentration~$\Gamma$.
For most simple liquids, the surface tension is a decreasing function of both variables but
the equation of state $\gamma (T,\Gamma)$ is in general neither linear in the concentration nor in the temperature. 
Let us denote $T_0$ and $\Gamma_0$ the equilibrium values. In the dilute limit, the surfactant monolayer reduces the interfacial tension according to~\cite{kralchevsky2009}
\begin{equation}
\gamma (T,\Gamma) = \gamma_0(T) -  \Gamma k_B T \ ,
\label{eqstate1}
\end{equation}
with $k_B$ the Boltzmann constant.
We also assume a linearized relation for the surface tension of the pure interface~\cite{birikhbook}
\begin{equation}
\gamma_0(T) = \gamma_0 -\gamma_T (T-T_0)  \ ,
\label{eqstate2}
\end{equation}
with $\gamma_0=\gamma_0(T_0)$  the equilibrium tension of the pure interface. This relation involves  the (positive) coefficient $\gamma_T = \vert \partial \gamma / \partial T \vert$ that characterizes the rate of change of  surface tension with respect to temperature. Experimental values for various liquids at room temperature are typically  of the order of $\gamma_T\sim 10^{-4}$~N$\cdot$m$\cdot$K$^{-1}$. The linear approximation eq.~(\ref{eqstate2}) is thus valid  in a wide range of temperatures around  $T_0 \sim 300$~K.
Coming back to the Marangoni boundary condition~(\ref{bcmarangoni}), the surface tension gradient then reads
\begin{equation}
\bm{\nabla}_{\parallel} \gamma = - \gamma_T  \bm{\nabla}_{\parallel} T -k_B T_0 \bm{\nabla}_{\parallel} \Gamma  \ ,
\label{gradient}
\end{equation}
where negligible cross-terms have been disregarded. 

The convective sweeping of the surfactants gives rise to restoring (solutal) Marangoni forces which oppose the thermocapillary flow. This mechanism thus provides an elastic feature to the interface, characterized by the coefficient $E_0 = \Gamma_0 k_B T_0$. It is also convenient to define the dimensionless elasticity number $\mathcal{E}$ as the ratio of solutal and thermal contributions to  Marangoni stresses~\cite{homsyJFM1984,carpenterJFM1985}
\begin{equation}
\mathcal{E} = \frac{\Gamma_0 \vert \partial \gamma/\partial \Gamma\vert}{ \Delta T \vert \partial \gamma / \partial T \vert } =  \frac{E_0}{\gamma_T \Delta T}  \ ,
\end{equation}
with $\Delta T$ the maximum temperature increase --- see the specific instances below. This dimensionless number characterizes the ``stiffening'' of the interface due to the presence of surfactant.

\subsection{Simplifying assumptions}

The dual Marangoni problem defined above  is non-linear and far too complex to be handled analytically. 
Some simplifications are therefore needed in order to be predictive. 

First, we restrict the discussion to the stationary regime. Although time-dependent behaviors might be relevant, regarding for instance the relaxation dynamics when the  source is swiched on or off~\cite{bickelSM2019}, they are not considered here. We also focus on radially symmetric sources and set $\mathbf{r}=(r,z)$, with $r=\sqrt{x^2+y^2}$. 

Second, we make the hypothesis that temperature variations  are sufficiently small so that all the parameters of the model  remain constant. The only temperature-depen\-dent quantity is the surface tension~$\gamma$ according to the equations of state~(\ref{eqstate1}) and~(\ref{eqstate2}).

Third, emphasis is placed on microscopic (\textit{i.e.}, submillimeter) length and velocity scales. We thus make the  hypothesis that both the Reynolds and the thermal P\'eclet numbers are small
\begin{equation}
\text{Re} = \frac{Ul}{\nu} \ll 1 \ , \quad \text{and} \quad \text{Pe}_{th}= \frac{Ul}{\chi} \ll 1 \ ,
\end{equation}
with $l$ and $U$ the characteristic length and velocity scales, and $\nu=\mu/\rho$ the kinematic viscosity.  In other words, the nonlinear contributions  in eqs.~(\ref{navierstokes}) and~(\ref{heatequation}) can be disregarded. Unlike its thermal counterpart, the solutal P\'eclet number 
\begin{equation}
\text{Pe}_{s}= \frac{Ul}{D} \ ,
\end{equation}
can take significant values. In water, the transport coefficients are of the order of $\nu \sim 10^{-6}$~m$^{2}\cdot$s$^{-1}$, $\chi \sim 10^{-7}$~m$^{2}\cdot$s$^{-1}$ and $D \sim 10^{-9}-10^{-10}$~m$^{2}\cdot$s$^{-1}$. The solutal P\'eclet number is thus expected to be 2 to 4 orders of magnitude larger than both other numbers.
The advection term in eq.~(\ref{diffusion}) therefore happens to be particularly relevant and will be the focus of  this work.

\section{Point-like source and deep water limit}
\label{deep}

We first discuss a spherical heat source located at the free interface of a semi-infinite liquid ($H\to \infty$). Assuming that the radius~$a$ of the source is small compared to $R$, heat may be regarded as emerging from a point source located at the origin. The power density is then written as $q(\mathbf{r})=Q \delta(\mathbf{r})$,  with $Q$ the total injected power. 

From a theoretical viewpoint, the structure of the thermocapillary flow due to a point-source is fairly well understood. In particular, the full non-linear  problem  defined by eqs.~(\ref{navierstokes})--(\ref{bcflux})  admits  an analytical solution in steady-state, which relies on the self-similar structure of the  fields~\cite{bratukhin1967}. 
Still, we focus on the regime of low Reynolds and thermal P\'eclet numbers $\text{Re} \ll 1$ and $\text{Pe}_{th}\ll 1$ in order to keep the calculations tractable. Indeed, the coupling with surfactant transport implies an additional non-linear contribution. In this section, the solutal P\'eclet number is supposed to be infinitely large: $\text{Pe}_s \gg 1$.

\subsection{Thermocapillary flow}

In the absence of surfactants, the flow is solely due to the thermocapillary effect. The temperature gradient gives rise to  surface stresses that puts the liquid in motion. 
The flow is directed from the heat source toward the edges of the cell.
Neglecting the advection term in the heat equation~(\ref{heatequation}),
the steady-state temperature can be written as
\begin{equation}
T(\mathbf{r})  = T_0 +\Delta T \frac{a}{\sqrt{r^2+z^2}} \ ,
\label{tempbulk}
\end{equation}
where we define  $a \Delta T \doteq Q/(2\pi \kappa)$. 
Even though there is no intrinsic length-scale in the model, this relation provides a formal link between the injected power  $Q$, the actual radius~$a$ of the heat source,  and the temperature rise $\Delta T$ at its surface.

The resulting flow field $\mathbf{v}^{(0)}(\mathbf{r})$ is then obtained as the solution of the linearized version of eqs.~(\ref{navierstokes})--(\ref{flowbc}). This classical issue is discussed for instance in ref.~\cite{wurgerJFM2014}. In particular, it is found that the radial component of the  interfacial velocity   is  simply given  by
\begin{equation}
v_r^{(0)}(r,0) = U \frac{a}{r}  \ .
\label{v0}
\end{equation}
Here, the thermocapillary velocity~$U$ is defined according to $a U \doteq \gamma_T Q/(4 \pi \kappa \eta)$.
The latter can also be expressed  as
\begin{equation}
U = \frac{\gamma_T  \Delta T}{2 \eta} \ .
\label{umarangoni}
\end{equation}
For typical values $ \gamma_T = 10^{-4}$~N$\cdot$m$\cdot$K$^{-1}$, $\eta =10^{-3}$~Pa$\cdot$s, and $\Delta T=2$~K, one expects the thermocapillary velocity to be of the order of   $U=10$~mm$\cdot$s$^{-1}$.

\subsection{Solutal counterflow}

When the interface is  covered with insoluble surfactants, both the temperature and the concentration gradients contribute to the Marangoni flow. In the Stokes regime $\text{Re} \ll 1$, the total velocity field can be written as the superposition of two terms: $\mathbf{v}(\mathbf{r})=\mathbf{v}^{(0)}(\mathbf{r})+\mathbf{v}^{(1)}(\mathbf{r})$. 
The first term~$\mathbf{v}^{(0)}$ is the thermocapillary contribution discussed above. The second contribution $\mathbf{v}^{(1)}$ describes the counterflow due to inhomogeneities of surfactant concentration.  Since the solutal P\'eclet number $\text{Pe}_s$ is large, the surface concentration is expected to be strongly coupled to the flow. The complexity thus arises from the non-linear advection term in eq.~(\ref{diffusion}).

We consider in this section the regime of infinite solutal P\'eclet number $\text{Pe}_s \to \infty$, where advection completely takes over diffusion. Since the mass flux vanishes at the boundaries,  the stationary axisymmetrical solution  of the transport eq.~(\ref{diffusion}) satisfies the simple relation
\begin{equation}
 v_r(r,0)  \Gamma (r) = 0 \ .
 \label{carpenter}
\end{equation}
This relation is the starting point of the seminal work of Carpenter and Homsy~\cite{carpenterJFM1985}. The latter was restricted to 2D geometry in the regime of shallow water, for which the lubrication approximation applies. Here we extend the analysis to 3D axisymmetric configuration in the deep water limit. We can deduce from eqn.~(\ref{carpenter}) that there  exists a critical radius $r_c$ such that
\begin{subequations}
\label{condbc}
\begin{align}
& \Gamma (r) = 0 \ , & 0\leq r <r_c \ , \label{concinf} \\
& v_r(r,0) =0 \ , & r > r_c   \ .  \label{vsup}
\end{align}
\end{subequations}
The first relation states that surfactant molecules are completely washed out from the depletion zone $r<r_c$. In the region $r>r_c$, the solutal counterflown~$\mathbf{v}^{(1)}$ exactly cancels the thermocapillary flow~$\mathbf{v}^{(0)}$ at the interface, so that the total velocity vanishes. 

The solutal contribution $\mathbf{v}^{(1)}$ is then sought as a regular solution of the Stokes equation.
To proceed, it is convenient to switch to  the (2D) Fourier representation. The problem being radially symmetric,  we introduce the Hankel transforms of order $\nu$~\cite{piessensbook}
\begin{align*}
&  \mathcal{H}_{\nu} \big[f(r) \big] = \tilde{f}(q) = \int_0^{\infty} r f(r)  J_{\nu} (qr) \d r    \ ,  \\
&  \mathcal{H}^{-1}_{\nu} \big[\tilde{f}(q) \big] = f(r) = \int_0^{\infty} q \tilde{f}(q)  J_{\nu} (qr) \d q    \ .  
\end{align*}
The main properties of Hankel transforms are summarized in appendix~\ref{appA}.
For convenience,  a different order $\nu$ is taken for the axial  $\tilde{v}^{(1)}_z (q,z)= \mathcal{H}_0 \big[v_z^{(1)} \big]$  and 
radial $\tilde{v}^{(1)}_r (q,z)= \mathcal{H}_1 \big[v_r^{(1)} \big]$ components of the velocity field~\cite{chraibiPoF2012}.
It can then be shown that $\tilde{v}^{(1)}_z$ satisfies a 4$^{th}$-order differential equation~\cite{bickelPRE2007}
\begin{equation*}
\left( \partial_z^4 -2q^2 \partial_z^2 + q^4 \partial_z^4 \right) \tilde{v}^{(1)}_z = 0 \ .
\end{equation*}
The solution that vanishes both at the interface $z=0$ and when \mbox{$z \to - \infty$} is thus 
\begin{equation}
\tilde{v}^{(1)}_z (q,z) = A(q) z e^{qz} \ ,
\label{vz1}
\end{equation}
where the amplitude $A(q)$ remains to be determined.
The radial component of the velocity can  be inferred from the  incompressibility condition~(\ref{incomp})  expressed in Hankel representation 
\begin{equation*}
q \tilde{v}^{(1)}_r + \partial_z  \tilde{v}^{(1)}_z =0 \ ,
\end{equation*}
so that we get
\begin{equation}
\tilde{v}^{(1)}_r(q,z)  = - q^{-1} A(q) (1+qz) e^{qz}  \ .
\label{vr1}
\end{equation}

\subsection{Concentration of surfactants}

The amplitude $A(q)$ is finally set by the balance of tangential stresses at the interface eq.~(\ref{bcmarangoni}), with the surface tension gradient specified in eq.~(\ref{gradient}). 
Since the thermocapillary contribution  is already supported by $\mathbf{v}^{(0)}$, the Marangoni boundary condition~(\ref{bcmarangoni}) leads to
\begin{equation*}
\eta \partial_z v^{(1)}_r (r,0) = -  \frac{E_0}{\Gamma_0} \partial_r \delta\Gamma \ ,
\label{mbc}
\end{equation*}
where we set $\delta\Gamma \doteq \Gamma - \Gamma_0$. Switching to Hankel representation, with $\delta \tilde{\Gamma} (q)=\mathcal{H}_0 \left[ \delta \Gamma  \right]$,
the latter relation leads to
\begin{equation}
 \delta \tilde{\Gamma} (q) = - \frac{2\eta \Gamma_0}{E_0} q^{-1}A(q) \ .
 \label{bchankel}
\end{equation}
We can now rewrite the conditional eqs.~(\ref{concinf}) and~(\ref{vsup}) in terms of Fourier integrals involving the unknown amplitude $A(q)$
\begin{subequations}
\label{dualbc}
\begin{align}
& \int_0^{\infty} A(q) J_0(qr)  \d q = \frac{E_0}{2\eta} \ , & 0\leq r <r_c \ , \label{eqd1} \\
& \int_0^{\infty}  A(q) J_1(qr)\d q   =U  \frac{a}{r}\ , & r > r_c   \ .  \label{eqd2}
\end{align}
\end{subequations}

The set of eqs.~(\ref{eqd1}) and~(\ref{eqd2}) belongs to the general class of dual integral equations, for which advanced mathematical methods are available~\cite{sneddonbook,duffybook}. Since the algebra is quite tedious but not essential for the comprehension, we skip the calculations (see app.~\ref{appB}) and focus on the analytical solution
\begin{equation}
A(q) = \frac{E_0}{\pi \eta}  \frac{\sin(qr_c)}{q} + \left( aU-\frac{r_cE_0}{\pi \eta} \right) \cos(qr_c) \ .
\label{solAq}
\end{equation}
Evaluation of the inverse Hankel transform is then straightforward. It can first be checked that $\delta \Gamma (r) = - \Gamma_0$ for $0 \leq r <r_c$, as expected. More interestingly,  one obtains for \mbox{$r>r_c$}
\begin{equation*}
\frac{\delta \Gamma (r)}{\Gamma_0}  =    -  \frac{2}{\pi} \arcsin \left( \frac{r_c}{r} \right) +  \frac{2}{\pi}\left( 1 - \frac{\pi \eta a U}{r_c E_0} \right) \frac{r_c}{\sqrt{r^2-r_c^2}}\ .
\end{equation*}
Clearly the last term diverges when $r\to r_c$. But remember that the actual value of $r_c$ has not been set yet.
Regularization of the latter expression then implies that the radius of the depletion zone is given by
\begin{equation}
r_c =  \frac{\pi \eta a U}{E_0}  =  \left(\frac{\pi}{2 \mathcal{E}}\right) a \ ,
\label{criticalradius}
\end{equation}
so that the concentration profile finally reads
\begin{equation}
\begin{cases}
\Gamma(r) = 0 \ , \qquad \text{for} \quad  0 \leq r < r_c \ , \\
\Gamma(r) = \Gamma_0 \displaystyle{\frac{2}{\pi}} \arccos \left( \frac{r_c}{r} \right) \ , \qquad \text{for}  \quad r>r_c \ .
\end{cases}
\end{equation}

\begin{figure}
\centering
\includegraphics[width=0.9\columnwidth]{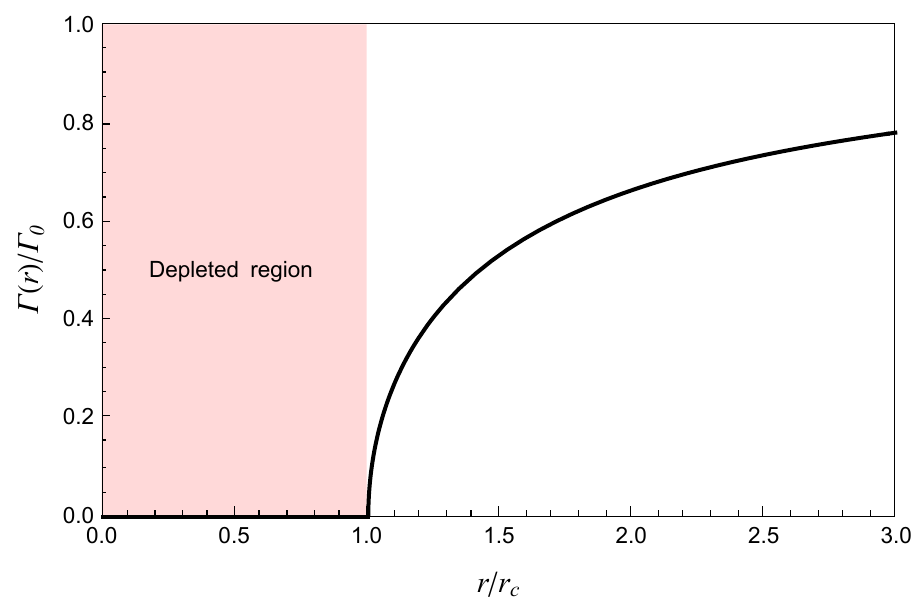}
\caption{Distribution of surfactants as a function of distance to the heat source, in the deep water limit and for $\text{Pe}_s \gg 1$. }
\label{gamma_deep}       
\end{figure}

We have thus completely determine  the distribution of surfactants --- see  fig.~\ref{gamma_deep}.
A particular feature of this solution is that 
 the concentration is singular at the border of the depletion zone $r \to r_c^+  $ 
\begin{equation}
\Gamma(r) \sim \left\vert r-r_c\right\vert^{\alpha} \ , 
\label{gammarc}
\end{equation}
with $\alpha = 1/2$.
Far away from the depletion zone $r \gg r_c$, the bare value $\Gamma(r) \to \Gamma_0$ is recovered~\cite{rmk}.
As expected, the size of the depletion zone increases linearly with the heating power~$Q\propto \Delta T$. It is also proportional to the coefficient $\gamma_T $
that set the velocity scale of the thermocapillary flow. Conversely, the effect of the counterflow becomes more and more pronounced as the surface elasticity 
increases. In the incompressible limit $\mathcal{E} \to \infty$, the stagnant zone eventually extends everywhere and the interface is completely frozen.

Regarding the velocity field, it is  obtained by evaluating the inverse Hankel transforms of eqs.~(\ref{vz1}) and~(\ref{vr1}). This operation has to be done numerically, excepted for the interfacial velocity $v_r(r,0)$. It is found to be strictly zero for $r>r_c$, as required from~(\ref{vsup}). For $0<r<r_c$ we get
\begin{equation}
v_r(r,0) = U \frac{a}{r} \sqrt{1 -   \left(  r /r_c\right)^2    } \ .
\end{equation}
We thus find that the velocity is hardly perturbed in a region near the origin, where $v_r(r,0) \sim_0 Ua/r$ [compare with eq.~(\ref{v0})].   At the border of the depleted region $r\to r_c^-$, the velocity exhibits a singular behavior as well
\begin{equation}
v_r(r,0) \sim  \left\vert r-r_c\right\vert^{\beta} \ , 
\label{vrrc}
\end{equation}
with the same exponent $\beta = 1/2$.

\section{Thermocapillary flow in a thin liquid film at finite solutal P\'eclet number}
\label{shallow}

The analysis developed in the previous section reveals that the variation of $\Gamma(r)$ is infinitely sharp at the border of the stagnant region.
Yet this singular behavior is expected to be smoothed when molecular diffusion is accounted for. Unfortunately, the regime of finite P\'eclet number is not easy to handle analytically for an unbounded liquid. To overcome this difficulty,  we consider the regime of shallow water  for which the lubrication approximation applies.

We focus hereafter on the thermocapillary flow in a thin liquid film with a free interface.  
In order to derive general and quantitative results regarding the coupling between the flow and the elastic response of the interface, we assume that the size~$R$ of the vessel is infinite.
The relevant length scale in this geometry is thus provided by the film thickness~$H$. 
The flow is driven by a temperature profile of the form 
\begin{equation}
T(\mathbf{r}) = T_0 + \Delta T \varphi(r)  \ , \label{tempgen}
\end{equation}
with $\Delta T >0$ a constant. The dimensionless function $\varphi(r)$ is assumed to be analytical and positive, and is such that $\varphi(0)=1$ and $\lim_{r\to \infty}\varphi(r)=0$. To keep the discussion as general as possible,  the functional form of $\varphi(r)$ will not be specified until sect.~\ref{illustration}. Still, we make the hypothesis that the width $\sigma$ of the temperature profile is much larger than the depth of the fluid.
We then characterize the re-organization of surfactant molecules in response to the thermocapillary flow, with particular emphasis on the regime of finite solutal P\'eclet number $\text{Pe}_s = HU/D$. In this section, the thermocapillary velocity is defined  as $U= \gamma_T \Delta T / (4 \eta)$.

\subsection{Marangoni flow in shallow water}

In the thin-film limit $H\ll \sigma$, the lubrication approximation applies and  the velocity $\mathbf{v}=v_r(r,z)\mathbf{e}_r$ follows the Stokes equation
\begin{equation}
\eta \partial_z^2v_r = \partial_r p  \ , \label{stokes}
\end{equation}
with  $p(r)$ the pressure.
This equation is solved together with the  boundary conditions~(\ref{flowbc}) at the interface, as well as the no-slip boundary condition  
at the bottom of the container: $\mathbf{v}(r,-H)=0$. The condition of vanishing flow rate in the stationary regime then provides an additional relation that eventually leads to
\begin{equation*}
v_r(r,z) = \frac{1}{4\eta H} \left(3 z^2+4zH+H^2 \right) \partial_r \gamma \ .
\end{equation*}
This expression states in particular that the interfacial velocity is directly proportional to the gradient of surface tension
\begin{equation}
v_r(r,0) = \frac{H}{4\eta} \partial_r \gamma  \ ,
\label{vshallowint}
\end{equation}
where $\partial_r \gamma$ depends on the inhomogeneities of both the temperature and the concentration fields.

\subsection{Infinitely large solutal P\'eclet number}

We first focus on the regime of large solutal P\'eclet number $\text{Pe}_s \gg 1$, where molecular diffusion can be neglected. The transport eq.~(\ref{diffusion}) then simplifies to  $v_r(r,0)  \Gamma (r) = 0$, so that the solution of Carpenter and Homsy applies~\cite{carpenterJFM1985}: $\Gamma(r) = 0$ for $r < r_c$, and $v_r(r,0)=0$ for $r>r_c$. According to eq.~(\ref{vshallowint}), the latter condition also implies that $\partial_r \gamma = 0$, and therefore
\begin{equation*}
\partial_r \Gamma = - \frac{\gamma_T \Gamma_0}{E_0} \partial_r T \ ,
\end{equation*}
 for $r>r_c$.
This equation is readily integrated for the generic temperature profile given by eq.~(\ref{tempgen}).  Enforcing both conditions $\Gamma(r_c)=0$
and $\lim_{r \to \infty} \Gamma(r) = \Gamma_0$, we end up with the general expression of the concentration profile for $r > r_c$
\begin{equation}
\Gamma(r) = \Gamma_0 \left( 1 -   \frac{\varphi(r)}{\varphi(r_c)}   \right) \ .
\label{concshallow}
\end{equation}
The radius $r_c$ of the depletion zone is obtained as the solution of the equation
\begin{equation}
 \varphi(r_c) = \mathcal{E}  \ ,
\label{rcshallow}
\end{equation}
with $\mathcal{E}$ the elasticity number.
Note that this solution only exists at ``low'' surface elasticity $\mathcal{E} \leq \max \left[ \varphi(r) \right]$. As soon as $\mathcal{E} > \max \left[ \varphi(r) \right]$, the thermocapillary flow is not strong enough to fully deplete the surfactants and the concentration remains finite. In this large elasticity regime, both contributions to the flow exactly cancels each other at $z=0$ and the interface stays immobile.

Although  the temperature profile has not been specified yet, 
one can still deduce from eq.~(\ref{concshallow})  that $\Gamma(r)$ is singular at the border of the depletion zone. Indeed,  when $r \to r_c^+$, the concentration behaves as
\begin{equation}
\Gamma(r) \sim \left\vert r-r_c \right\vert^{\alpha} \ ,
\end{equation}
with $\alpha =1$, whatever the functional form of $\varphi(r)$. Note that this value differs from that obtained in  deep water, where it was found $\alpha = 1/2$.
Confinement therefore has a drastic effect on the behavior of $\Gamma(r)$ in the vicinity of~$r_c$.

\subsection{Finite solutal P\'eclet number}

The previous discussion is now extended to the regime of finite solutal P\'eclet number. In stationary state, advective and diffusive currents balance each other and the mass transport eq.~(\ref{diffusion}) leads to
\begin{equation*}
D \partial_r \Gamma = v_r(r,0) \Gamma(r) \ .
\end{equation*}
Inserting expression~(\ref{vshallowint}) for the interfacial velocity and taking into account the equation of state~(\ref{gradient}), we arrive  at
\begin{equation*}
\partial_r \Gamma = - \text{Pe}_s \Gamma(r) \partial_r  \left( \mathcal{E} \frac{\Gamma(r)}{\Gamma_0}  +   \varphi(r)  \right) \ .
\label{eqc}
\end{equation*}
Although nonlinear, this differential equation admits an analytical solution. Together with the boundary condition at infinity, $\lim_{r \to \infty} \Gamma(r) = \Gamma_0$, one finds 
\begin{equation}
\ln \frac{\Gamma(r)}{\Gamma_0} +\mathcal{E}\text{Pe}_s \left( \frac{\Gamma(r)}{\Gamma_0}-1 \right) = -  \text{Pe}_s \varphi(r) \ .
\label{cimplicit}
\end{equation}
This implicit relation can be discussed in the relevant limits. 
Let us first assume that $\mathcal{E}$ is constant. Taking the limit $\text{Pe}_s \to \infty$, one recovers the concentration profile eq.~(\ref{concshallow}) previously derived, whereas a constant value $\Gamma(r)=\Gamma_0$ is predicted when $\text{Pe}_s \to 0$.
We next consider a fixed value of the P\'eclet number. In the low-elasticity regime $\mathcal{E}\ll1$,  the concentration behaves as 
\begin{equation}
\Gamma(r) \sim \Gamma_0 \exp\left[ - \text{Pe}_s \varphi(r) \right] \ .
\end{equation}
We thus expect a strong depletion of surfactants  in the vicinity of the central axis.
In the opposite regime of large elasticity number $\mathcal{E}\gg1$, the concentration returns rapidly to its original value $\Gamma_0$.

\begin{figure}
\centering
\includegraphics[width=\columnwidth]{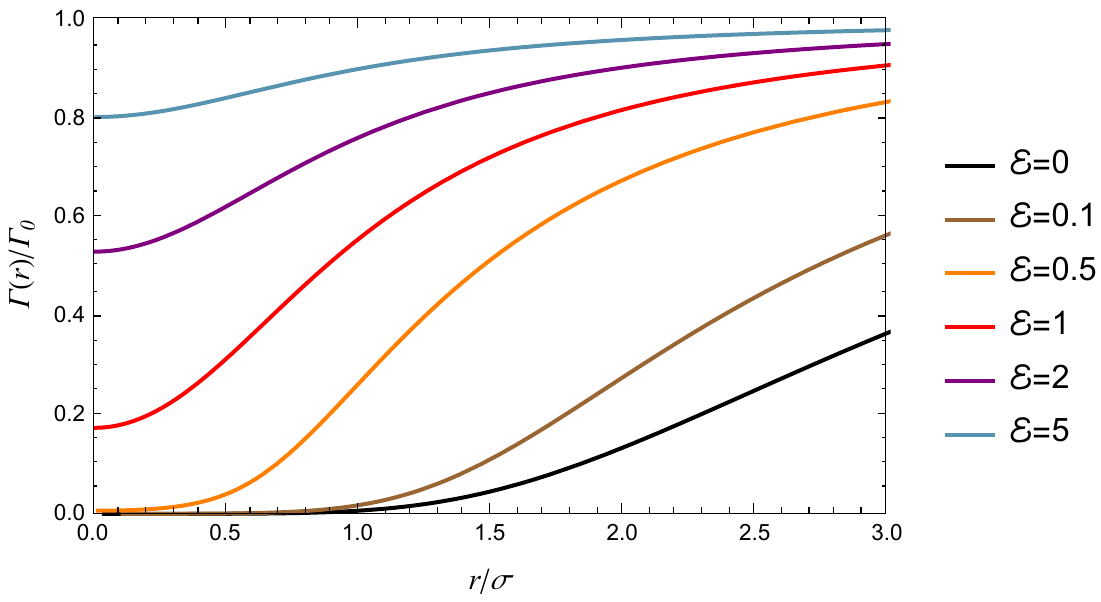}
\caption{Concentration of surfactants at fixed P\'eclet number $\text{Pe}_s=10$, for different values of the elasticity number $\mathcal{E}$.}
\label{gamma_shallow}       
\end{figure}

\subsection{Thermocapillary flow due to laser absorption}
\label{illustration}

As an illustration, we consider a thin liquid layer  heated by a cylindrical laser beam that propagates normal to the interface. Several  experimental and numerical studies~\cite{corderoPRE2009,rivierePRE2016}  recently reported that the resulting temperature field in the liquid phase can be very well fitted by a Lorentzian profile  
\begin{equation}
T(\mathbf{r}) = T_0 + \frac{\Delta T}{ 1+ \displaystyle{r^2/\sigma^2}} \ ,
\label{templor}
\end{equation}
with  $\Delta T$ the maximum temperature increase along the $z$-axis. It is also observed that the half-width of the profile $\sigma$ is significantly larger that the original beam radius~\cite{corderoPRE2009,rivierePRE2016}. 

Let us first discuss the regime of large P\'eclet number $\text{Pe}_s \gg 1$. Solving eq.~(\ref{rcshallow}) with the Lorentzian profile eq.~(\ref{templor}), it is straightforward to get the  critical radius 
\begin{equation}
r_c = \sigma \sqrt{\mathcal{E}^{-1} - 1} \ .
\end{equation}
As mentioned earlier, this solution only exists at low surface elasticity $\mathcal{E} < 1$. As soon as $\mathcal{E} \geq 1$, the solutal counterflow exactly compensate the thermocapillary flow and the interface is completely motionless. 

For  finite values of the P\'eclet number, the concentration profile is obtained by inserting eq.~(\ref{templor}) in the general solution eq.~(\ref{cimplicit}).  The implicit relation can then be inverted numerically.
The result is shown on fig.~\ref{gamma_shallow}  for $\text{Pe}_s=10$.
As the elasticity number increases, the extension of the depletion zone first decreases until the point where it totally vanishes. At larger elasticity numbers,   
surfactants are still depleted around the origin but the concentration remains finite everywhere. Eventually, the concentration becomes uniform  when $\mathcal{E}\to \infty$.
The main conclusion of this section is that the singularities of $\Gamma(r)$ predicted in the limit $\text{Pe}_s \to \infty$ actually disappear as soon as $\text{Pe}_s$ is finite.

\section{Conclusion}
\label{conclusion}

To summarize, we have developed a theoretical analysis of the flow properties when both thermal and solutal Marangoni effects are relevant.
The flow is primarily driven by thermal effects, but the elastic response of the interface brings an additional complexity that was first discussed in ref.~\cite{carpenterJFM1985}. 
Our main results are the following:
\begin{itemize}
\item in the deep water limit and for the convective regime of mass transport ($\text{Pe}_s \gg 1$), surfactant molecules are swept away from a region whose size is inversely proportional to the elasticity number. Outside this region, the solutocapillary and thermocapillary contributions to the interfacial flow exactly cancel each other, even though the interfacial stress does not vanishes.
\item if the elasticity of the interface is too large, the depleted region may even disappear and the concentration remains finite everywhere. This corresponds to the situation where the stagnant region extends over the entire interface.
\item the effect of diffusive mass transport  is discussed in the shallow water regime. For finite $\text{Pe}_s$, the transition between the depleted and the stagnant regions of the interface becomes smooth.
\end{itemize}

Some simplifications were necessary in order to derive our analytical results. In particular, it is assumed all along the article that the water-air interface remains flat. But it is well known that, under confinement, Marangoni stresses also drive interfacial deformations. According to refs.~\cite{robertPRF2016,kavokineAngew2016,leitePRF2018}, the deformation amplitude $\delta h$ is expected to scale as $\delta h \sim \gamma_T \Delta T / (\rho g H)$. For typical values $ \gamma_T = 10^{-4}$~N$\cdot$m$\cdot$K$^{-1}$, $\Delta T=1$~K, and for $H=10^{-3}$~m, the relative height is $\delta h/H \sim 10^{-2}$: deformations can thus be safely neglected for a liquid layer whose thickness lies in the millimeter range. Nevertheless, the assumption breaks down for smaller thicknesses, in which case a steady-state solution may even cease to exist as the liquid film would eventually no longer wet the substrate~\cite{crasterRMP2009}. Conversely, thermogravity effects might become relevant for larger thicknesses, but taking into account these additional contributions is beyond the scope of this study.

Although the contamination of the water-air interface is a long-standing issue in interfacial science,  quantitative experimental data are rather sparse. The competition between a divergent outward flow and the solutal inward response may thus be used as practical way to evidence the presence of impurities. This requires to refine the theoretical models in order to provide reliable predictions regarding observable quantities. The results presented in this work is one attempt in this direction, and would definitely deserved to be tested experimentally.

\appendix

\section{Some properties of Hankel transforms}
\label{appA}

For the sake of completeness, we remind in this appendix the main properties of the Hankel transforms. The reader is referred to ref.~\cite{piessensbook} for a detailed presentation at an introductory level. 

Let $f(r)$ be a function defined for $r \geq 0$, and such that $f(r) \to 0 $ when $r \to \infty$.  The $\nu$th order ($\nu \in \mathbb{N}^*$) Hankel transform of  $f(r)$ is then defined as 
\begin{equation*}
\mathcal{H}_{\nu} \big[f(r) \big] = \tilde{f}(q) = \int_0^{\infty} r f(r)  J_{\nu} (qr) \d r    \ ,  \\
\end{equation*}
and the inversion formula reads
\begin{equation*}
\mathcal{H}^{-1}_{\nu} \big[\tilde{f}(q) \big] = f(r) = \int_0^{\infty} q \tilde{f}(q)  J_{\nu} (qr) \d q    \ .  
\end{equation*}
Because there is no simple linearization formula for Bessel functions, Hankel transforms do not have as many elementary properties as Fourier transforms do. Still, they are a powerful tool in order to solve problems with cylindrical symmetry. Define for instance the differential operator
\begin{equation*}
\Delta_{\nu} \doteq  \frac{\d^2}{\d r^2} + \frac{1}{r} \frac{\d }{\d r} - \left( \frac{\nu}{r} \right)^2   \ ,
\end{equation*}
then we have
\begin{equation*}
\mathcal{H}_{\nu} \big[ \Delta_{\nu} f(r) \big] = -q^2  \mathcal{H}_{\nu} \big[f(r) \big]  \ .
\end{equation*}
Regarding the derivative of a function, the Hankel transforms satisfy the following couple of relations
\begin{equation*}
\mathcal{H}_{\nu} \left[  r^{-\nu-1} \frac{\d }{\d r} \Big( r^{\nu+1} f(r) \Big) \right] = q  \mathcal{H}_{\nu+1} \big[f(r) \big]  \ ,
\label{nup1}
\end{equation*}
and
\begin{equation*}
\mathcal{H}_{\nu} \left[  r^{\nu-1} \frac{\d }{\d r} \Big( r^{1-\nu} f(r) \Big) \right] = -q  \mathcal{H}_{\nu-1} \big[f(r) \big]  \ .
\label{num1}
\end{equation*}
The latter relation may be used for instance to derive eq.~(\ref{bchankel}) in the text. For $\nu=1$ and $f(r)=\delta \Gamma(r)$, one indeed gets
$\mathcal{H}_{1}[\partial_r \delta \Gamma(r)]=-q \mathcal{H}_{0}[\delta \Gamma(r)]$, and the result follows straightforwardly.

\section{Solution of the integral eqs.~(\ref{dualbc})}
\label{appB}

In this appendix, we simply check that the expression of $A(q)$ given in the text [eq.~(\ref{solAq})] does satisfy the dual set of integral eqs.~(\ref{dualbc}). 
The full derivation of the solution is more involved, and is based on a general method that is  discussed in details elsewhere~\cite{bickelPreprint2019}. 

Let us first recall some general properties of integrals involving Bessel functions. For $0 \leq r < t$, one has
\begin{align*}
& I_1(r,t) = \int_0^{\infty}  q^{-1} \sin (qt) J_0(qr)  \d q = \frac{\pi}{2} \ , \\
& I_2(r,t) =  \int_0^{\infty}  \cos (qt)J_0(qr) \d q = 0 \ , 
\end{align*}
whereas for $r  > t$ one gets
\begin{align*}
& I_3(r,t) = \int_0^{\infty}  q^{-1} \sin (qt) J_1(qr) \d q = \frac{t}{r} \ , \\
& I_4(r,t) = \int_0^{\infty} J_1(qr) \cos (qt) \d q = \frac{1}{r} \ .
\end{align*}
It is then straightforward to verify that, for $r<r_c$, one obtains
\begin{align*}
\int_0^{\infty} A(q) & J_0(qr)  \d q   \\
&  = \frac{E_0}{\pi \eta} I_1(r,r_c) + \left( aU-\frac{r_cE_0}{\pi \eta} \right) I_2(r,r_c)  = \frac{E_0}{2 \eta} \ .
\end{align*}
For $r >r_c$, it is found
\begin{align*}
\int_0^{\infty} A(q) & J_1(qr)  \d q   \\
&  = \frac{E_0}{\pi \eta} I_3(r,r_c) + \left( aU-\frac{r_cE_0}{\pi \eta} \right) I_4(r,r_c)  = \frac{aU}{r} \ .
\end{align*}
As a consequence, both integral relations, namely eqs.~(\ref{eqd1}) and~(\ref{eqd2}), are satisfied.


\begin{thebibliography}{}
\bibitem{rmk}
Strictly speaking, the asymptotic value of the concentration should be $\Gamma_0 + \delta \Gamma_0$, where $\delta \Gamma_0$ is set by mass conservation. It can actually be shown that $\delta \Gamma_0 \sim 1/R$, with $R$ the size of the container. The correction is thus irrelevant in the limit $R \to \infty$.
\end{thebibliography}
\end{document}